\documentclass[preprint,showpacs,showkeys,preprintnumbers,amsmath,amssymb]{revtex4}

\usepackage{graphicx}
\usepackage{dcolumn}
\usepackage{bm}

\begin{document}

\title{Valence fluctuation in Ce$_{2}$Co$_{3}$Ge$_{5}$ and crystal field effect in Pr$_{2}$Co$_{3}$Ge$_{5}$}
\author{Samar Layek}
\author{V. K. Anand}
\email{vivekkranand@gmail.com}
\author{Z. Hossain}
\affiliation{Department of Physics, Indian Institute of Technology, Kanpur 208016, India}
\begin{abstract}

Polycrystalline samples of ternary rare earth germanides R$_{2}$Co$_{3}$Ge$_{5}$(R = La, Ce and Pr) have been prepared and investigated by means of magnetic susceptibility, isothermal magnetization, electrical resistivity and specific heat measurements. All these compounds crystallize in orthorhombic U$_{2}$Co$_{3}$Si$_{5}$ structure (space group \textit{Ibam}). No evidence of magnetic or superconducting transition is observed in any of these compounds down to 2 K. The unit cell volume of Ce$_{2}$Co$_{3}$Ge$_{5}$ deviates from the expected lanthanide contraction, indicating non trivalent state of Ce ions in this compound. The reduced value of effective moment ($\mu_{eff}$ $\approx$ 0.95 $\mu_{B}$) compared to that expected for trivalent Ce ions further supports valence fluctuating nature of Ce in Ce$_{2}$Co$_{3}$Ge$_{5}$. The observed temperature dependence of magnetic susceptibility is consistent with ionic interconfiguration fluctuation (ICF) model. Although no sharp anomaly due to a phase transition is seen, a broad Schottky-type anomaly is observed in the magnetic part of specific heat of Pr$_{2}$Co$_{3}$Ge$_{5}$. An analysis of $C_{mag}$ data suggests a singlet ground state in Pr$_{2}$Co$_{3}$Ge$_{5}$ separated from the singlet first excited state by 22 K and a doublet second excited state at 73 K.

\end{abstract}

\pacs{71.28.+d, 65.40.Ba, 71.70.Ch}
\keywords{Intermetallic compounds, Magnetization, Specific heat, Valence fluctuation, Crystal field effect}
\maketitle


\section*{Introduction}

Cerium and praseodymium based intermettalics are well known for their diverse physical properties and have been investigated by many research groups. The ground state of the cerium compounds is determined by the relative strength of Ruderman-Kittel-Kasuya-Yosida (RKKY) and Kondo interactions. While Kondo effect has a tendency to lead to a nonmagnetic ground state, RKKY interaction tries to establish long range magnetic order. The two characteristic energy scales associated with these interactions are Kondo temperature $T_{K} \sim exp(-1/\mid JN(E_F)\mid))$ and $T_{RKKY} \sim \mid JN(E_F)\mid ^{2}$ where \textit{J} is the coupling constant and $N(E_{F})$ is the density of states at the Fermi level. Competition between the RKKY interaction and Kondo effect is best described by the well known Doniach phase diagram \cite{1}. Exotic phenomena like heavy fermion behaviour, non-Fermi liquid behaviour and quantum criticality are observed in the regime where these two characteristic energy scales become comparable. Magnetic order is observed in the regime where RKKY interaction dominates over the Kondo effect, whereas nonmagnetic valence fluctuating behaviour is observed in Kondo interaction dominant regime. Since in valence fluctuating systems the valence of rare earth ions keeps on fluctuating between 4$f^{n}$ and 4$f^{n-1}$ states, the physical properties of such systems get significantly modified. Valence fluctuation in Ce-based compounds is of special interest and has been observed in many systems, such as CePd$_{3}$ \cite{2}, CeNiIn \cite{3}, CeRhIn \cite{4} and CeNi$_{4}$B \cite{5}. In contrast, the physical properties of Pr-compounds are strongly influenced by crystal field effect. The nine fold degenerate ground states of Pr (\textit{J} = 4) split into a combination of CEF-split states. An interesting consequence of the crystal filed effect is the realization of excitonic mass enhancement leading to heavy fermion behavior as in PrOs$_{4}$Sb$_{12}$ \cite{6,7} and PrRh$_{2}$B$_{2}$C \cite{8}.

Ternary compounds R$_{2}$T$_{3}$X$_{5 }$ (where R = Ce and Pr, T = Transition metal and X = Si and Ge) having orthorhombic U$_{2}$Co$_{3}$Si$_{5}$ structure (space group \textit{Ibam}) show interesting properties like Kondo lattice behavior, heavy fermion behavior, magnetic ordering and valence fluctuation. For example, literature show pressure induced superconductivity in Kondo antiferromagnet Ce$_{2}$Ni$_{3}$Ge$_{5}$ \cite{9,10}, valence fluctuation in Ce$_{2}$Ni$_{3}$Si$_{5}$ \cite{11} and anomalous magnetoresistance in R$_{2}$Ni$_{3}$Si$_{5}$ (R\textit{ }= Pr, Dy, Ho) \cite{12}. Ce$_{2}$Pd$_{3}$Si$_{5}$ \cite{13}, Ce$_{2}$Rh$_{3}$Ge$_{5}$ and Ce$_{2}$Ir$_{3}$Ge$_{5}$ \cite{14} are known antiferromagnetic Kondo lattice system with moderate heavy-fermion behavior in the last two compounds. The Pr-based compounds, Pr$_{2}$Ni$_{3}$Ge$_{5}$ \cite{15}, Pr$_{2}$Ni$_{3}$Si$_{5}$ \cite{16} and Pr$_{2}$Pd$_{3}$Ge$_{5}$ \cite{17} order magnetically, and Pr$_{2}$Rh$_{3}$Ge$_{5}$ exhibit moderate heavy fermion behavior \cite{17}. In our effort to search for interesting compounds of 235 composition we have synthesized and investigated R$_{2}$Co$_{3}$Ge$_{5}$ (R = La, Ce and Pr). In this paper we report our results of electrical, magnetic and thermal properties on these compounds.

\section*{Experimental}

The polycrystalline samples of R$_{2}$Co$_{3}$Ge$_{5}$ (R = La, Ce and Pr) have been prepared by standard arc melting technique. Appropriate stoichiometric amount of high purity elements (La 99.99\%, Pr 99.99\%, Ce 99.999\%, Co 99.9995\% and Ge 99.9999\%) were arc melted on a water cooled copper hearth under argon atmosphere. During the melting process ingots were flipped and remelted several times for homogenizing. The weight loss for each sample during melting was less than 0.5\%. The melted as-cast polycrystalline samples were then sealed in an evacuated quartz tube and annealed at 1000 $^{0}$C for a week to improve the phase purity.  Crystal structure and phase purity were checked by Cu K$_\alpha$ X-ray diffractometer. To check the composition and homogeneity we used scanning electron microscope (SEM) equipped with energy dispersive X-ray analysis (EDAX). We used a commercial superconducting quantum interference device (SQUID) magnetometer for magnetization measurement. Specific heat was measured using the relaxation method in a physical properties measurement system (PPMS-Quantum Design). The electrical resistivity was measured by the standard \textit{ac} four-probe techniques using the \textit{AC}\textit{-}transport measurement option of the PPMS.

\section*{Results and discussions}

The powder X-ray diffraction data and scanning electron microscope (SEM) images reveal the single phase nature of R$_{2}$Co$_{3}$Ge$_{5}$ (R = La, Ce, Pr) samples. They crystallize in U$_{2}$Co$_{3}$Si$_{5}$-type orthorhombic structure (space group \textit{Ibam}) with lattice parameters and unit cell volumes as shown in Table I. The EDAX analysis confirmed the desired stoichiometry of 2:3:5 for all these compounds. The impurity phase(s) are estimated to be less than 5 \% in La$_{2}$Co$_{3}$Ge$_{5}$ and Pr$_{2}$Co$_{3}$Ge$_{5}$, and about 8\% in Ce$_{2}$Co$_{3}$Ge$_{5}$. We notice that the unit cell volumes of R$_{2}$Co$_{3}$Ge$_{5}$ (R = La, Ce, Pr) do not follow the trend of lanthanide contraction. The unit cell volume of Ce$_{2}$Co$_{3}$Ge$_{5}$ significantly deviates from the expected lanthanide contraction behaviour suggesting that Ce ions in this compound are not in Ce$^{3+}$ state, rather they are in mixed-valence state which is further supported by the low value of effective moment. Similar kind of behavior has been found in CeRhSb \cite{18}. Further, a valence fluctuation behaviour in elemental Ce is accompanied with a large volume collapse at the $\gamma$-$\alpha$ transition between two isostructural fcc phases of Ce. Despite the intensive study for last eight decades starting from the work by P. Bridgman \cite{19} the intriguing phenomena that occurs at $\gamma$-$\alpha$ transition is still not completely understood \cite{20,21,22}. While in $\gamma$-phase 4\textit{f}-electrons are in localized state (Ce$^{3+}$), in $\alpha$-phase they are delocalized (Ce$^{4+}$). The volume collapse occurs due to partial delocalization of localized \textit{f} moments and the hybridization of \textit{f}-band and valence band of Ce.

\begin{table}
\caption{\label{tab:table1} Lattice parameters and unit cell volumes of R$_{2}$Co$_{3}$Ge$_{5}$ (R = La, Ce and Pr)}
\begin{ruledtabular}
\begin{tabular}{lcccc}
 Compounds & a (\AA) & b (\AA) & c (\AA) & V (\AA$^3$) \\

\hline
La$_{2}$Co$_{3}$Ge$_{5}$ & 9.883 & 11.983 & 5.943 & 703.2 \\

Ce$_{2}$Co$_{3}$Ge$_{5}$ & 9.802 & 11.777 & 5.941 & 684.8 \\

Pr$_{2}$Co$_{3}$Ge$_{5}$ & 9.798 & 11.760 & 5.951 & 685.4 \\
\end{tabular}
\end{ruledtabular}
\end{table}

\section*{A. Ce$_{2}$Co$_{3}$Ge$_{5}$}

The magnetic susceptibility data of Ce$_{2}$Co$_{3}$Ge$_{5}$ exhibit paramagnetic behaviour down to 2 K. The inverse magnetic susceptibility $\chi^{-1} (T)$ of Ce$_{2}$Co$_{3}$Ge$_{5}$ deviates significantly from the expected Curie-Weiss behaviour below 140 K (Fig. 1); such a shape of inverse susceptibility curve is typical for valence fluctuating Ce-compounds. A reasonable fit is obtained for susceptibility data to the modified Curie-Weiss behavior, $\chi (T) = \chi_{0} + C/(T- \theta_{p})$, in the temperature range 30 -- 300 K with $\chi_{0}$ = 1.44 $\times$ 10$^{-3}$ emu/Ce-mole, Curie constant $C$ = 0.11258, and $\theta_{p}$ = -3.05 K. Effective magnetic moment, $\mu_{eff}$ = 0.95 $\mu_{B}$/Ce obtained from the Curie constant is very low compared to the magnetic moment of free Ce$^{3+ }$ions (2.54 $\mu_{B}$) but higher than that of non-magnetic Ce$^{4+}$ state (0 $\mu_{B}$). The intermediate value of the effective magnetic moment in Ce$_{2}$Co$_{3}$Ge$_{5}$ may be due to the mixed-valence state of the Ce ions. Such kind of behavior has been found in CeNi$_{4}$B with effective magnetic moment value 0.52 $\mu_{B}$ \cite{5} and 0.45 $\mu_{B}$ \cite{23} by two different research groups. The inset of Fig. 2 shows the magnetic field dependence of isothermal magnetization $M(B)$ of Ce$_{2}$Co$_{3}$Ge$_{5 }$ at 5 K. As expected, magnetization is almost linear with field up to the investigated field of 3 T, however the magnitude is very small consistent with mixed-valence behaviour.

\begin{figure}
\includegraphics[width=8.5cm, keepaspectratio]{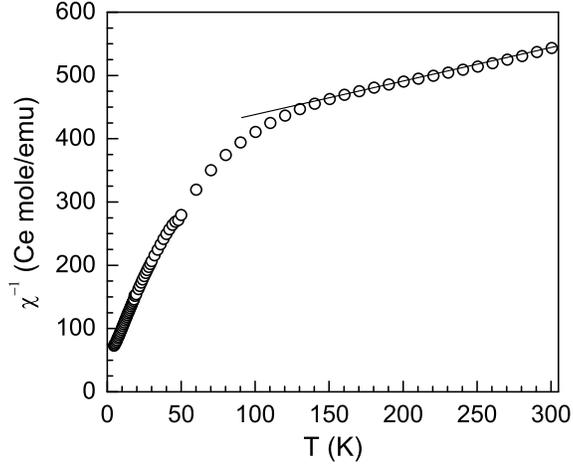}
\caption{\label{fig1} The temperature dependence of inverse magnetic susceptibility, $\chi^{-1} (T)$ of Ce$_{2}$Co$_{3}$Ge$_{5}$ at 1 T; solid line is a guide to eye.}
\end{figure}

\begin{figure}
\includegraphics[width=8.5cm, keepaspectratio]{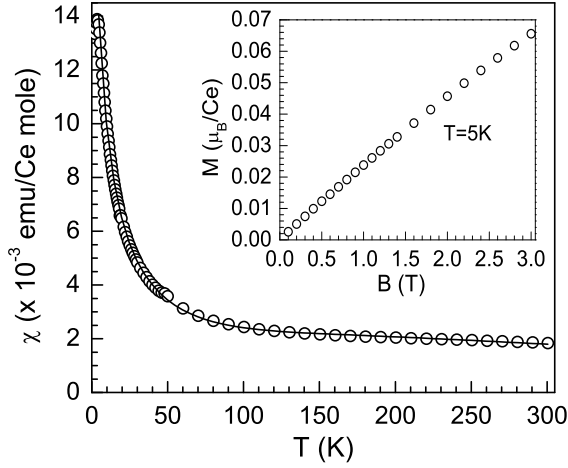}
\caption{\label{fig2} Magnetic susceptibility $\chi (T)$ as a function of temperature for Ce$_{2}$Co$_{3}$Ge$_{5}$ measured at 1 T. The solid line shows the fit to two-level ionic interconfiguration fluctuations (ICF) model as discussed in text. Inset shows the isothermal magnetization $M(B)$ as a function of magnetic field at 5 K.}
\end{figure}

The above results clearly indicate a valence fluctuating behaviour in Ce$_{2}$Co$_{3}$Ge$_{5}$. We therefore made an attempt to interpret the magnetic susceptibility data with the two-level ionic interconfiguration fluctuations (ICF) model. The theory of ICF was first proposed by Hirst \cite{24,25} and latter Sales and Wohlleben applied it to explain the valence fluctuating susceptibility behaviour observed in few Yb systems \cite{26}. In ICF model the observed physical quantity is calculated by taking average over the two ionic configurations 4$f^{n}$ and 4$f^{n-1}$ which compete for stability. Following Sales and Wohlleben \cite{26}, and Franze et al. \cite{27}, the modified ICF susceptibility is given by

\begin{displaymath}
 \chi (T)= \frac{N \{ \mu_n^2 \nu (T) + \mu_{n-1}^2 [1 - \nu (T)]\} }{3 k_B T^*}
\end{displaymath}

with

\begin{displaymath}
 \nu (T)= \frac{(2 J_n + 1)}{(2 J_n + 1) + (2 J_{n-1} + 1) exp(- \frac{E_{ex}}{k_B T^*})}
\end{displaymath}

and

\begin{displaymath}
 T^*= \sqrt{T^2+T_{sf}^2}
\end{displaymath}

\noindent where $ \nu (T)$ is the fractional occupation of ground state, $\mu_{n}$ and $\mu_{n-1}$ are the effective moments in 4$f^{n}$ and 4$f^{n-1}$ states, and $(2J_{n}+1)$ and $(2J_{n-1}+1)$ are the degeneracies of the corresponding energy states $E_{n}$ and $E_{n-1}$, interconfigurational excitation energy $E_{ex} = E_{n} - E_{n-1}$, and $T_{sf}$ is the spin fluctuation temperature that characterizes the valence fluctuation; $T_{sf}= \hbar \omega_{f}/k_{B}$, $\omega_{f}$ being the rate of fluctuation between the two valence states 4$f^{n}$ and 4$f^{n-1}$. If we take Ce$^{4+}$ (\textit{J} = 0 and $\mu$ = 0) state as ground state and Ce$^{3+ }$ (\textit{J} = 5/2 and $\mu$ = 2.54 $\mu_{B}$) as excited state, interconfiguration fluctuations between 4$f^{1}$ and 4$f^{0}$ states of Ce lead to

\begin{displaymath}
 \chi (T)= \frac{N (2.54 \mu_B)^2 [1 - \nu (T)]}{3 k_B T^*}
\end{displaymath}

and

\begin{displaymath}
 \nu (T)= \frac{1}{1 + 6 exp(- \frac{E_{ex}}{k_B T^*})}
\end{displaymath}

In order to take care of the contribution from the stable Ce$^{3+}$ ions which may be present as impurity in Ce$_{2}$Co$_{3}$Ge$_{5}$, we add a term \textit{$\chi$}\textit{$_{imp}$} = \textit{$\chi$}\textit{$_{0}$} + \textit{nC}/(\textit{T-$\theta$}) to the magnetic susceptibility and get

\begin{displaymath}
 \chi (T)= (1-n) \frac{N (2.54 \mu_B)^2 [1 - \nu (T)]}{3 k_B T^*} + n \frac{C}{T-\theta} + \chi_0
\end{displaymath}

A fit of the magnetic susceptibility data of Ce$_{2}$Co$_{3}$Ge$_{5}$ to this equation shows a nice agreement in the experimentally observed data and the ionic ICF model (solid line in Fig. 2) with the fraction of stable Ce$^{3+}$ ions \textit{n} = 0.172, $\theta$ = -7.52 K and $\chi_{0}$ = 1.5 x 10$^{-4}$ emu/Ce mole, and $E_{ex}$ = 495 $\pm$ 24 K and $T_{sf}$ = 108 $\pm$ 8 K. These values of $E_{ex}$ and $T_{sf}$ are in the range of typical values found for the valence fluctuating systems, as for example, for Ce$_{2}$Ni$_{3}$Si$_{5}$, $E_{ex}$ = 285 K and $T_{sf}$ = 140 K \cite{11}, and for CeRhSb, $E_{ex}$ = 368 K and $T_{sf}$ = 87 K \cite{18}.

\begin{figure}
\includegraphics[width=8.5cm, keepaspectratio]{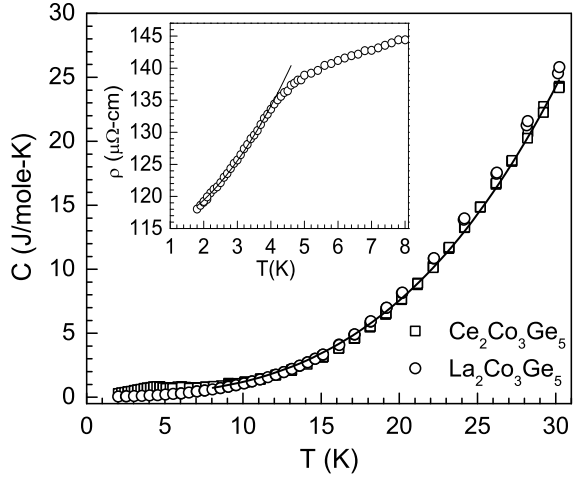}
\caption{\label{fig3} Temperature dependence of the specific heat of Ce$_{2}$Co$_{3}$Ge$_{5 }$ and La$_{2}$Co$_{3}$Ge$_{5 }$ in the temperature range 2--30 K; solid line shows the fit to $C (T) = \gamma T + \beta T^{3}$. Inset shows the low temperature electrical resistivity, $\rho (T)$ of Ce$_{2}$Co$_{3}$Ge$_{5}$ in the temperature range 2--8 K; solid line is the fit to $\rho (T) = \rho(0) + A T^{2}$.}
\end{figure}

Fig. 3 shows the temperature dependence of the specific heat $C (T)$ of Ce$_{2}$Co$_{3}$Ge$_{5}$ and non-magnetic analog La$_{2}$Co$_{3}$Ge$_{5}$ in the temperature range 2 -- 30 K. The specific heats of Ce$_{2}$Co$_{3}$Ge$_{5}$ and La$_{2}$Co$_{3}$Ge$_{5 }$ are comparable over the wide temperature range and do not show any pronounced anomaly. A weak anomaly in the specific heat data of Ce$_{2}$Co$_{3}$Ge$_{5}$ around 4.5 K is attributed to unidentified impurity phase(s) as observed in XRD and SEM results. On fitting the specific heat data of Ce$_{2}$Co$_{3}$Ge$_{5}$ with the expression $C (T) = \gamma T + \beta T^{3}$ in the temperature range 8 -- 30 K (solid line in Fig. 3) we got Sommerfeld coefficient $\gamma$ = 17 mJ/Ce mole K$^{2}$. From a linear fit of $C (T) vs. T^{2}$ graph below 10 K the specific heat coefficient \textit{$\gamma$} is estimated to be $\sim$ 10 mJ/La mole K$^{2}$ for La$_{2}$Co$_{3}$Ge$_{5}$.

The inset of Fig. 3 shows the electrical resistivity $\rho (T)$ of Ce$_{2}$Co$_{3}$Ge$_{5}$ in the temperature range 2 -- 8 K. Despite the presence of weak anomaly around 4.5 K (which we attribute to the impurity phase) low temperature resistivity data are reasonably consistent with the Fermi-liquid behaviour typically observed in valence fluctuating systems \cite{27}. Solid line in the inset represents the fit to the expression $\rho (T) = \rho(0) + A T^{2}$ with $\rho (0)$ = 114 $\mu \Omega$ cm, and \textit{A} = 1.22 $\mu \Omega$ cm/K$^{2}$.

\begin{figure}
\includegraphics[width=8.5cm, keepaspectratio]{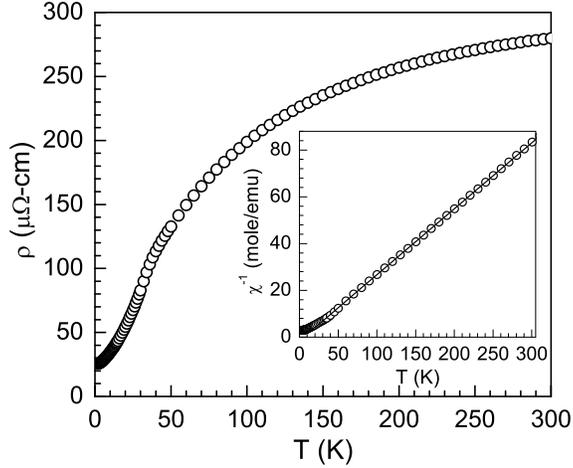}
\caption{\label{fig4} Electrical resistivity $\rho (T)$ of Pr$_{2}$Co$_{3}$Ge$_{5}$ in the temperature range 2--300 K. Inset shows the inverse magnetic susceptibility data as a function of temperature measured at 1 T; solid line is the fit to the Curie-Weiss behaviour.}
\end{figure}

\section*{B. Pr$_{2}$Co$_{3}$Ge$_{5}$}

The magnetic susceptibility data of Pr$_{2}$Co$_{3}$Ge$_{5}$ show that it remains paramagnetic down to 2 K. The temperature dependence of inverse magnetic susceptibility $\chi^{-1} (T)$ of Pr$_{2}$Co$_{3}$Ge$_{5}$ is shown in the inset of Fig. 4 for an applied field of 1 T. Above 50 K the inverse susceptibility data is consistent with Curie-Weiss law, $\chi (T) = C/(T- \theta_{p})$. A linear fit of inverse susceptibility gives effective magnetic moment $\mu_{eff}$ = 3.76 $\mu_{B}$/Pr, and paramagnetic Curie temperature $\theta_{p}$ = 5.48 K. The effective magnetic moment is little higher than the value expected for free Pr$^{3+}$ ions (3.58 $\mu_{B}$). Deviation from the Curie-Weiss behavior below 50 K may be attributed to crystal field effect. The isothermal magnetization (data not shown) exhibit linear field dependence as expected for a paramagnetic system.

The temperature dependence of the zero-field electrical resistivity $\rho (T)$ of Pr$_{2}$Co$_{3}$Ge$_{5}$ is shown in Fig. 4. Resistivity decreases almost linearly with decreasing temperature down to 200 K but deviates significantly below this. The broad curvature around 80 K may be the result of crystal field effect. High value of residual resistivity ratio (RRR), $\rho(300K)/\rho(2K)$ = 11 and a low residual resistivity of 25.6 $\mu \Omega$-cm (at 2 K) indicate good quality of our polycrystalline Pr$_{2}$Co$_{3}$Ge$_{5}$ sample.

\begin{figure}
\includegraphics[width=8.5cm, keepaspectratio]{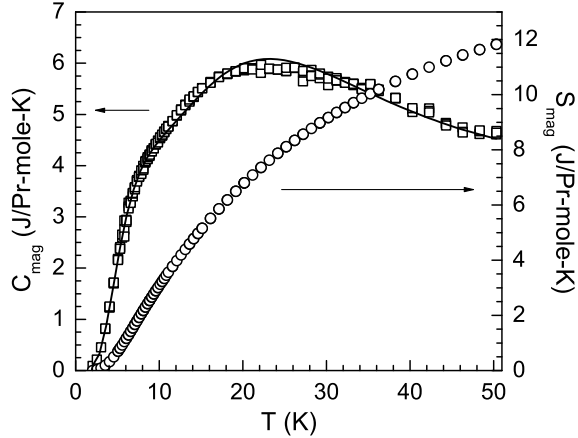}
\caption{\label{fig5} Magnetic part of the specific heat and entropy as a function of temperature in the temperature range 2--50 K. The solid line shows the fit to three level crystal field scheme as discussed in the text.}
\end{figure}

Specific heat of Pr$_{2}$Co$_{3}$Ge$_{5}$ does not show any pronounced anomaly down to 2 K excluding the existence of magnetic or superconducting transition. However, a broad Schottky type anomaly is seen near 10 K which is more clear in the magnetic contribution to the specific heat $C_{mag} (T)$ of Pr$_{2}$Co$_{3}$Ge$_{5 }$ (Fig. 5) which we obtained after subtracting the lattice contribution from the specific heat of Pr$_{2}$Co$_{3}$Ge$_{5}$ assuming it to be roughly equal to that of non magnetic analog La$_{2}$Co$_{3}$Ge$_{5}$. The experimentally observed \textit{C}$_{mag}$ data could be reproduced reasonably with three level crystal electric field scheme. Schottky contribution to the specific heat from three low lying crystal field levels is given by

\begin{displaymath}
 C_{Sch}(T)= \frac{R \{ g_0 g_1 (\frac{\Delta_1}{T})^2 exp(- \frac{\Delta_1}{T}) + g_0 g_2 (\frac{\Delta_2}{T})^2 exp(- \frac{\Delta_2}{T}) + g_1 g_2 (\frac{\Delta_1 - \Delta_2}{T})^2 exp(- \frac{\Delta_1 + \Delta_2}{T}) \} }{ \{ g_0 + g_1 exp(- \frac{\Delta_1}{T}) +g_2 exp(- \frac{\Delta_2}{T}) \}^2 }
\end{displaymath}

\noindent where $\Delta_{i}$ are the splitting energies with respect to ground state, and $g_{i}$ are the corresponding degeneracy of the levels. Solid line in Fig. 5 represents the fit to this expression with the splitting energies $\Delta_{1}$ = 22 K and $\Delta_{2}$ = 73 K, and $g_{0}$ = 1, $g_{1}$ = 1 and $g_{2}$ = 2, i.e., Pr$_{2}$Co$_{3}$Ge$_{5}$ has a singlet ground state separated from a singlet first excited state by 22 K and a doublet second excited state at 73 K. That the ground state is a singlet lying below a singlet excited state at 22 K is further supported by the temperature dependence of magnetic entropy which attains a value of $Rln(2)$ at 17 K. Presence of crystal field effect makes it difficult to estimate the value of Sommerfeld coefficient \textit{$\gamma$} precisely for Pr$_{2}$Co$_{3}$Ge$_{5}$. However, at 2 K $C/T$ has a value of 57 mJ/Pr mole K$^{2}$; an extrapolation of $C/T$ {\it vs}. $T^{2}$ plot below 5 K gives a rough estimate of $\gamma$ $\sim$ 20 mJ/Pr mole K$^{2}$.

\section*{Summary and conclusions}

We have synthesized and investigated the magnetic and transport properties of two new rare earth intermetallic compounds Ce$_{2}$Co$_{3}$Ge$_{5}$ and Pr$_{2}$Co$_{3}$Ge$_{5}$. Both of these compounds have paramagnetic ground state. Valence fluctuating behaviour is inferred from the departure of unit cell volume of Ce$_{2}$Co$_{3}$Ge$_{5}$ from the expected lanthanide contraction. The observed magnetic susceptibility behaviour is well represented by ionic interconfiguration fluctuation (ICF) model confirming the valence fluctuating behaviour in Ce$_{2}$Co$_{3}$Ge$_{5}$. The magnetic and transport properties of Pr$_{2}$Co$_{3}$Ge$_{5}$ are strongly influenced by the crystal field effect. Absence of magnetic order in Pr$_{2}$Co$_{3}$Ge$_{5}$is attributed to CEF-split nonmagnetic singlet ground.

\section*{Acknowledgement}

We acknowledge CSR Indore for providing access to low temperature measurements using PPMS.

\end{document}